\documentclass{iaus}
\usepackage{graphicx}

\newcommand{\aap}{A\&A}
\newcommand{\apj}{ApJ}
\newcommand{\apjl}{ApJ}

\newcommand{\mnras}{MNRAS}
\newcommand{\Mstar}{M_\star}
\newcommand{\Msun}{M_\odot}
\newcommand{\Mcl}{M_{\rm cl}}
\newcommand{\nat}{Nature}

\title[Feedback and Molecular Cloud Turbulence] 
{Theory of Feedback in Clusters and Molecular Cloud Turbulence}

\author[V\'azquez-Semadeni]   
{Enrique V\'azquez-Semadeni}

\affiliation{Centro de Radioastronom\'\i a y Astrof\'\i sica, UNAM,
Campus Morelia \\ P.O. Box 3-72 (Xangari), Morelia, Michoac\'an,
M\'exico \\ email: {\tt e.vazquez@.crya.unam.mx}}

\pubyear{2011}
\volume{270}  
\pagerange{}
\jname{Computational Star Formation}
\editors{J. Alves, B.G. Elmegreen, J.M. Girart, and V. Trimble, eds.}
\begin{document}

\maketitle

\begin{abstract}
I review recent numerical and analytical work on the feedback from both
low- and high-mass cluster stars into their gasoeus environment.
The main conclusions are that i) outflow driving appears capable of
maintaing the turbulence in parsec-sized clumps and retarding their
collapse from the free-fall rate, although there exist regions within
molecular clouds, and even some examples of whole clouds, which are not
actively forming stars, yet are just as turbulent, so that a more
universal turbulence-driving mechanism is needed; ii) outflow-driven
turbulence exhibits specific spectral features thatcan be tested
observationally; iii) feedback plays an important role in reducing the
SFR; iv) nevertheless, numerical simulations suggest feedback cannot
completely prevent a net contracting motion of clouds and
clumps. Therefore, an appealing source for driving the turbulence
everywhere in GMCs is the accretion from the environment, at all
scales. In this case, feedback's most important role may be to prevent a
fraction of the gas nearest to newly formed stars from actually reaching
them, thus reducing the SFE.

\keywords{ISM: clouds, methods: numerical, stars: winds, outflows; turbulence}
\end{abstract}

\firstsection 
\section{Introduction}

Stars form in dense cores within molecular clouds (MCs), and
simultaneously they feed back on their environment through a variety of
mechanisms, including ionizing radiation, winds, outflows, and supernova
explosions. This feedback crucially affects the stars' environment, in
particular its physical conditions, evolution, as well as the subsequent
pattern of star formation (SF). In particular, it is believed that
stellar feedback may be responsible for the observed non-thermal motions
in clouds and clumps, which are generally interpreted as small-scale
supersonic turbulence (\cite{ZE74}). The latter, in turn, is thought to
provide support against the clouds' self-gravity, and to be able to
maintain a quasi-virial equilibrium state in the clouds, thus preventing
global collapse and maintaining a low global SFR. This notion is at the
basis of several theoretical models of SF (e.g., \cite[McKee
1989]{McKee89}; \cite{Matzner02}; \cite{KM05}; \cite{HC08}). 
In this contribution I review
recent results on the role of feedback in driving the turbulence in MCs
and their substructure, specifically on its coupling with the ambient
gas, on the ability of the turbulence to support the clouds, and on its
role in regulating the SFR and the SF efficiency (SFE). For a discussion
on the role of feedback in the determination of the stellar masses, see
the review by Bate in this volume.

\section{Background} \label{sec:bckgnd}

More than half a century ago, \cite{Oort54} suggested that an
interstellar cycle should exist, in which the HII regions produced by
the ionizing radiation of O stars would form a dense shell which, upon
fragmenting, would produce cold cloudlets with a significant velocity
dispersion. These would then grow by coalescence until they became
gravitationally unstable, at which point they would form new stars, and
the cycle would repeat. This {\it Oort model} was later formulated
mathematically by \cite{FS65}, who
concluded that the SFR should be proportional to the square of the gas
density, in agreement with the SF law known at the time
(\cite{Schmidt59}), and that the cloud mass spectrum should scale as
$M^{-3/2}$, is in agreement with the observed mass
spectrum of CO clouds (see, e.g., \cite{Blitz93} and references therein)
and of low-mass clumps (\cite{MAN98}). These results exemplify the
fundamental implications of feedback on their environment.

Another seminal study was that of Norman \& Silk (1980, hereafter NS80),
who considered the low-mass analogy of the Oort-Field-Saslaw model, with
the driving energy provided in this case by the outflows produced by
low-mass T-Tauri stars. In their model, the collision of expanding,
wind-driven shells would form dense clumps that would then evolve under
the competition of growth by coalescence, and destruction by leak, drag,
or very energetic collisions. 


Based on the Oort-NS80 scenario, \cite{McKee89} advanced a model in
which MCs form from atomic cloudlets that grow by coalescence, and
become molecular (in the sense that most of the carbon is in CO) when
the extinction $A_V \sim 1$. He argued that at this point the clouds
must be magnetically supercritical, and therefore begin to contract as a
whole, although their substructures (clumps and cores) are still
subcritical, and are thus magnetically supported. Nevertheless, the
contraction speeds up ambipolar diffusion (AD) in the cores, rendering
them supercritical and allowing them to collapse and form stars. The
latter begin driving turbulence in the cloud, eventually halting the
collapse, and allowing the cloud to reach a stable equilibrium. This
occurs when the mean extinction has increased to $A_V \sim 4$--8. In
this model, the gas depletion time for typical giant MCs (GMCs) was
estimated to be $\sim 2$--$4 \times 10^8$ yr, and the typical magnetic
field strength $B \sim 20$--$40 \mu$G.

Today, however, we know that clouds do not grow only by coalescence, but
that instead a significant, and perhaps dominant, mechanism determining
their mass is direct accretion from their warm, diffuse surrounding
medium, allowing for much shorter growth timescales (\cite{BS80};
\cite{HP99}; Ballesteros-Paredes et al.\ 1999a,b; \cite{KI02};
\cite{AH05}; \cite{HA07}; \cite{VS_etal06}; \cite{Banerjee_etal09}), and
that this mechanism can drive strong turbulence in the clouds.  Thus,
more recent research has focused on the details of the physics that
regulates the energy transfer from the stellar sources, especially
bipolar outflows from low-mass stars, to their environment, as well as
to whether indeed stellar feedback is capable of feeding the clouds'
turbulence and maintaining them in rough hydrostatic equilibrium. I now
briefly summarize some of the main results in these respects.

\section{Feedback from outflows} \label{sec:outflows}

\subsection{Efficiency of coupling with the environment} \label{sec:coupling}

Estimates of the amount of energy injected to the environment by an
outflow ($\sim 10^{47}$ erg per $\Msun$; \cite{Shu_etal88}) suggest that
outflows may deposit enough kinetic energy in their parent clump as to
maintain its turbulence (see \cite{RB01} and references therein).
However, an important, recurrent question is whether the bipolar
outflows couple effeciently to their environment. 

In this regard, \cite{Quillen_etal05} investigated the correlation of
the molecular gas kinematics in NGC1333 with the distribution of young
stellar objects (YSOs) within this cloud, finding that the velocity
dispersion does not vary significantly across the cloud and is
uncorrelated with the number of nearby young stellar outflows. However,
they did find about 20 cavities in the velocity channel maps that they
interpreted as remnants of past outflow activity. Those authors
concluded that, while outflows may not directly drive the turbulence in
the clumps, the cavities (``fossil outflows'') may provide an efficient
coupling mechanism to the environment. Later numerical work by
\cite{Cunningham_etal06b} supported this conclusion.

The interaction of single, or a few, outflows with their environment has been
extensively studied. \cite{dCR05} performed magnetohydrodynamic (MHD)
simulations of high-density clumps propagating through a high-density
medium, in order to represent the interaction of a jet with the
medium. They found that $\sim 10$--30\% of the jet's kinetic energy can
be transferred to the medium, depending on the magnetic field's strength
and orientation relative to the jet.
\cite{Cunningham_etal06a} investigated the effect of collisions
among outflows, finding that the collisions reduce the efficiency of
transfer to the cloud, so that the most efficient drivers are isolated
outflows. However, \cite{Banerjee_etal07} noted, using isothermal
simulations, that outflows should be inefficient drivers of {\it
supersonic} turbulent motions in their parent clouds because the
compressions they produce can only re-expand sonically, although then
\cite{Cunningham_etal09} argued that this problem can be
circumvented if the cloud is previously turbulent (i.e, the turbulence
is {\it maintained} rather than generated), and adequate care is taken
of modeling the cooling. 

It can be concluded from this section that, through the mediation of
long-lived cavities, magnetic fields, and a pre-existent turbulent
velocity field, outflows appear capable of at least maintaining the
turbulent motions in a clump. 

\subsection{Feedback and nature of the turbulence}
\label{sec:turb_nature}

In real MCs, stars are almost always born in clusters, and so the effect
of an {\it ensemble} of outflows on their parent clump is also a crucial
issue. In particular, many studies have focused on the nature of
the turbulent motions induced in the parent clump by an ensemble of
outflows. We now turn to this issue.

Numerical simulations of evolving clumps at the parsec scale suggest
that the initial (``interstellar'') turbulence is quickly replaced by
``outflow'' turbulence, at least within the scales modeled by these
simulations (\cite{LN06}; \cite{NL07}). This transition consists in a
secular variation of the topology of the density and velocity fields in
the clump. The density field develops a central concentration, with a
power-law profile of the form $\rho \sim r^{-3/2}$, and the velocity
field develops a circulation pattern, with gravity driving infall
motions that balance the outward motions driven by the
outflows. Moreover, the turbulent energy spectrum develops a knee at a
characteristic ``outflow scale'' (\cite{Matzner07}), and a slope steeper
than that of isotropic random driving at scales smaller than this
(\cite{NL07}; Carroll et al.\ 2009, 2010).

A particularly interesting feature is that the presence of
the outflows does not seem to impede the development of coherent streams
of infalling gas towards the center of the gravitational potential well
(\cite{NL07}). This implies that the most massive stars forming in the
clump are fed from a mass supply that extends out to the scale of the
whole clump, instead of being restricted to the scale of the very dense
core that contains the forming star (\cite{Wang_etal10}). This result is
in stark contrast with the currently popular notion that the masses of
forming stars are determined by the masses of the cores in which they
reside (e.g., \cite{PN02}, \cite{KMK05}, \cite{Alves_etal07}), and more
in agreement with the scenario of competitive accretion for star
formation (e.g., \cite{Bonnell_etal97}; \cite{BB06}; \cite{Bonnell_etal07};
\cite{Smith_etal09}), in which the material reaching a star is collected
from distant locations (up to several tenths of a parsec away) within
the clump.  In fact, the development of a density profile with slope
$-3/2$ may also be indicative of a generalized state of collapse in the
clumps, as it is the signature of ongoing dynamic collapse
(\cite{Shu77}). 

In summary, the simulations discussed above suggest that outflow
feedback is capable of maintaining the turbulence within parsec-scale
clumps that are already forming stars. However, there exist a few
observational features of clouds that cannot be explained by this
mechanism. First, it is well known that the majority of a cloud's volume
is not in the process of forming stars; the star-forming regions within
clouds are generally limited to only a few localized, high-column
density spots (e.g, \cite{Kirk_etal06}) within the clouds. Moreover,
there exist clouds with very little or no significant star-forming
activity, such as Maddalena's cloud (\cite{MT85}) or the Pipe Nebula
(\cite{Onishi_etal99}; see also \cite{Lombardi_etal06}) that
nevertheless have turbulent properties essentially indistinguishable
from those of clouds that are actively forming stars. Outflow feedback
clearly cannot explain the origin or maintenance of the turbulence in
the non-star-forming regions.

Second, principal component analysis (PCA) of spectroscopical line
emission data (\cite{HeBr07}; \cite{Brunt_etal09}) shows that the
turbulent velocity dispersion in the clouds and their substructure
appears to be dominated by large-scale, dipolar, velocity gradients
spanning the entire structure. 
Such universal large-scale nature of the turbulent motions in clouds
does not seem to be attainable by outflows, due to their
small-scale, localized nature. Note, however, that Carroll et al.\
(2010) have recently questioned these results, a claim that requires
further investigation. In any case, the problem of driving the
turbulence in non-star-forming regions remains, and in general it can be
argued that a different source of turbulence is required there.

\section{Cloud evolution and control of the SFE}
\label{sec:life_SFE} 

\subsection{Cloud destruction} \label{sec:destruction}

The expansion of HII regions from massive stars, especially from those
that are sufficiently close to the cloud's periphery to produce a
``blister HII region'', is probably the dominant feedback mechanism at the
scale of GMCs (for a discussion of the contributing feedback mechanisms,
see \cite{Matzner02}), and is generally believed to be capable of
effectively dispersing the cloud within $10^7$ yr after SF starts in the
cloud (\cite{BS80}). \cite{Whitworth79} analytically estimated the
eroding effect of O stars producing blister HII regions in MCs, and
concluded that the cloud would be completely dispersed after only 4\% of
its mass had been converted to stars, assuming a standard
\cite{Salpeter55} initial mass function (IMF). 

Extending on this result, and considering that a fraction of the massive
stars are completely interior to the cloud and do not produce a blister
HII region, \cite{FST94} estimated the maximum number of OB stars that
can be hosted by a GMC without it being destroyed, concluding that the
resulting SFE should range between 2\% and 16\%, with an average of
5\%. On the observational side, \cite{WM97} compared the Galactic
distribution of GMC masses to the distribution of OB association
luminosities, in order to statistically estimate the GMC mass that
contains at least one O star. They found that the median GMC mass to
satisfy this condition is $\sim 10^5 \Msun$, and that the average SFE is
$\sim 5$\%. Also, they estimated that the typical GMC lifetime as the time
for the cloud to be photoevaporated by the O stars is
$\sim 3 \times 10^7$ yr, although a more recent estimate by
\cite[Matzner (2002)]{Matzner02} yields a somewhat shorter timescale of
$\sim 2 \times 10^7$ yr.

\subsection{Support of isolated clouds and the SFE} \label{sec:support}


Together with the erosion inflicted on clouds by massive-star feedback,
it is also generally believed that the latter can also maintain GMCs
close to virial equilibrium for times significantly longer than their
free-fall times. Thaking this as a working assumption, the SFR and SFE
can be derived, similarly to the procedure used in studies of SF
self-regulation due to outflows (e.g., \cite{FC83}; \cite[McKee
1989]{McKee89}). For example, \cite[Matzner (2002)]{Matzner02}
analytically derived a cloud destruction timescale of $\sim 2 \times
10^7 (M_{\rm cl}/10^6 \Msun)^{-1/3}$ yr, implying that more massive
clouds should be destroyed in shorter times.

A semi-analytic model of the energy balance in GMCs
was presented by \cite{KMM06}. In this model, the fully time-dependent
Virial Theorem was written and solved numerically for a spherical cloud
under the influence of its self-gravity and the HII-region feedback,
with no a-priori assumption of equilibrium. The
result was that clouds undergo a few expansion-contraction oscillations,
until they are finally dispersed, with lower-mass clouds ($M \sim 2
\times 10^5 \Msun$) are more quickly dispersed (typically within $\sim
1.5$ crossing times) than more massive ones ($M \sim 5 \times 10^6
\Msun$), which last $\sim 3$ crossing times. Note that this result is
opposite to that of \cite[Matzner (2002)]{Matzner02}. The SFEs
over the clouds' lifetimes were found to be $\sim 5$--10\%.

Full numerical simulations including self-consistent SF prescriptions,
aimed at studying the role of stellar feedback on reducing the SFR have
been generally carried out at the clump-scale level. There is a general
agreement that feedback reduces the SFR and, consequently, the SFE of a
cloud, regardless of whether the energy is injected by low-mass outflows
(\cite{NL07}; \cite{Wang_etal10}), low-mass protostellar luminosity
(\cite{Bate09}; \cite{PB09}; \cite{Offner_etal09}), or high-mass-star
winds (\cite{DB08}). Although the details vary depending on the specific
implementation, all studies conclude that the SFR is reduced with
respect to the case with no feedback. In particular, some of these
studies quantify the SFE, reporting values $\lesssim 10$\% per free-fall
time (e.g., \cite{Wang_etal10}), in agreement with observations of
similar regions (e.g., \cite{Evans_etal09}), although \cite[Dale \&
Bonnell (2008)]{DB08} warn that the the SFR may accelerate in time,
rendering any conclusions based on the average SFE uncertain.

\subsection{Evolution of cloud-environment systems}
\label{sec:cloud-env} 

All the numerical studies mentioned in the previous section have been
performed at the clump scale, thus omitting the interaction between the
clumps and the larger-scale cloud in which they are immersed. This may
be of crucial importance, since recent simulations of GMC formation and
evolution in the presence of self-gravity (\cite{VS_etal07}, 2009;
\cite{HB07}; \cite{HH08}; \cite{Heitsch_etal08};
\cite{Hennebelle_etal08}; \cite{Banerjee_etal09};
see also the review by \cite{VS10}) as well as observations of
massive-star forming regions (\cite{GM_etal09}; \cite{Csengeri_etal10})
suggest that there is a continuous infalling flow from the large to the
small scales, most probably driven by gravity. In this case, the mass
reservoir of the local star-forming regions is not limited to its
immediate environment, but rather includes material from regions farther
away in the cloud than the local clump.

A numerical study incorporating the effect of massive-star ionizing
radiation in the context of globally contracting clouds was recently
performed by \cite{VS_etal10}. These authors considered the evolution of
clouds formed by converging flows in the warm neutral atomic medium
(WNM), and followed it until the time of active SF.
In this type of simulations, the clouds are found to enter a global
state of contraction, causing the localized star-forming regions within
the clouds to have a continuous inflow of mass from their environment,
rather than having a fixed mass. Thus, the SFE, defined as SFE$ = \Mstar
/(\Mcl + \Mstar)$, where $\Mstar$ is the total mass in stars and $\Mcl$
is the gas mass of the cloud, can maintain realistic values, of order of
a few percent for GMCs, over extended periods of time, because the gas
mass is replenished by the infall while the cloud continues forming
stars. These authors also found that the feedback acts on size scales
much smaller than the gravitational potential well of whole GMC, and
therefore the global inflow persists, even if locally the gas on route
to forming stars is dispersed, reducing the SFE. This effect is stronger
for more massive clouds, which have deeper and more extended potential
wells. Smaller clouds were found to be more easily destroyed, in
agreement with the semi-analytic model of \cite{KMM06}. However, the
results from \cite{VS_etal10} suggest that perhaps termination of the SF
activity on the scale of the largest GMCs may require the termination of
the inflows, rather than being accomplished by the feedback. Further
exploration of parameter space and feedback modeling is needed in order
to obtain firmer conclusions in this regard.

The possibility that clouds are in a generalized state of contraction
and accreting from WNM has the additional advantage that it may provide
the needed universal source of turbulence in GMCs and their
substructure, since it is by now well established that the dense layers
produced by converging flows naturally develop turbulence, as a
consequence of several instabilities acting on them
(\cite{Hunter_etal86}; \cite{Vishniac94}; \cite{FW06};
\cite{Heitsch_etal06}; \cite{VS_etal06}). Recently, \cite{KH10} have
compared the energy input rate from the accretion to the energy
dissipation rate by the turbulence, concluding that the former is
sufficient to maintain the turbulence even if only 10\% of the accretion
energy is converted to turbulence in the dense regions. These authors
suggest that this mechanism can operate at all scales from Galactic
disks to protostellar disks, passing through GMCs and their
substructure, extending the suggestion that a universal mass cascade,
driven by gravity, exists at all scales within GMCs (\cite{FBK08}).

\section{Discussion and conclusions} \label{sec:conclusions}

The results from the works discussed in this review imply that feedback
from clusters has a complex and strongly nonlinear effect on their
parent clouds and clumps, which still remains elusive in some
respects. Numerical simulations of outflows from low-mass stars in
parsec-sized clumps agree in general that the outflows inject sufficient
momentum into the clump to sustain the turbulence in it. In these
simulations, the turbulence develops a peculiar form of the turbulent
energy spectrum, with a knee at the characteristic outflow scale, and a
steep slope ($\sim -2.5$) below that scale. These features should in
principle be observationally detectable, and so this prediction is
directly testable. However, we pointed out that outflow driving cannot
account for the turbulence in non-star-forming regions of clouds, or in
clouds that have no significant star-forming activity at all. Thus, we
concluded that a more universal source of turbulence is needed. 

Analytical and semi-analytical calculations of the effect of feedback
from massive stars at the GMC scale suggest that the feedback is also
capable of slowing down the SFR to realistic rates, while simultaneoulsy
supporting the cloud over a few crossing times, after which the cloud is
finally destroyed. However, by their very nature, these models cannot
account for the spatial distribution of the stellar sources, an
ingredient which is suggested to be crucial by numerical studies of the
formation and evolution of a large molecular complex. Such simulations
show that the entire cloud complex begins contracting gravitationally
even before it becomes predominantly molecular, so that, by the time a
GMC is fully formed and begins to form stars, it is already
contracting. Moreover, the contraction occurs in a highly non-uniform
fashion, due to the turbulence produced at the cloud's formation, so
that SF and their feedback occur at a few localized spots in the
clouds. This prevents the feedback from reaching the more distant, yet
also infalling, regions of the clouds.

This suggests again that the source of the turbulence at the scale of
whole GMCs should be a more universal one than the feedback, which is
applied very locally. Within this scenario, a natural candidate source
is the accretion energy from the environment, since it is well known
that the dense layers formed by colliding streams are naturally
turbulent due to several dynamical instabilities acting on them. In this
picture, the GMCs are the dense ``layers'' within converging WNM flows,
while clumps and cores are the dense ``layers'' within converging molecular
and/or cold atomic convergent flows, all probably driven by gravity
rather than by the stellar feedback. The latter is only a byproduct of
the gravitational contraction, and acts mainly to prevent a fraction of
the infalling gas mass from actually reaching the forming stars,
reducing the SFR and the SFE from their free-fall value. More
theoretical, numerical and observational work is clearly needed to
confirm this picture, and sort out its details.

\end{document}